\documentclass[12pt]{article}
\usepackage{graphicx}
\usepackage{bm}        
\usepackage{amssymb}   
\usepackage{hyphenat}

\input epsf
\def\beq{\begin{eqnarray}}
\def\eeq{\end{eqnarray}}

\def\kk{q}

\def\ba{\begin{eqnarray}}
\def\ea{\end{eqnarray}}
\def\beq{\begin{eqnarray}}
\def\eeq{\end{eqnarray}}

\def\mpl{M_{\rm Pl}}

\def\d{\mathrm{d}}
\def\p{{\cal P}}

\def\L*{{\cal L}_*}
\def\L{\mathcal{L}}
\def\({\left(}
\def\){\right)}
\def\ie{{\it i.e. }}
\def\nn{\nonumber}
\def\p{\partial}
\def\mn{_{\mu \nu}}

\def\p{\partial}

\def\<{\langle}
\def\>{\rangle}

\def\lsim{\mathrel{\rlap{\lower3pt\hbox{\hskip0pt$\sim$}}
     \raise1pt\hbox{$<$}}}         
\def\gsim{\mathrel{\rlap{\lower4pt\hbox{\hskip1pt$\sim$}}
     \raise1pt\hbox{$>$}}}         

\def\lsim{\mathrel{\rlap{\lower3pt\hbox{\hskip0pt$\sim$}}
     \raise1pt\hbox{$<$}}}         
\def\gsim{\mathrel{\rlap{\lower4pt\hbox{\hskip1pt$\sim$}}
     \raise1pt\hbox{$>$}}}         

\usepackage{amsmath}
\usepackage{amsfonts}
\usepackage{verbatim}

\begin{document}

\begin{titlepage}

\begin{flushright}
{NYU-TH-09/18/10}

\today
\end{flushright}
\vskip 0.9cm

\centerline{\Large \bf Cosmic Acceleration and the Helicity-0 Graviton}
\vskip 0.7cm
\centerline{\large Claudia de Rham$^a$, Gregory Gabadadze$^b$,}
\centerline{\large Lavinia Heisenberg$^a$\ and David Pirtskhalava$^b$}
\vskip 0.3cm

\centerline{\em $^a$D\'epartment de Physique  Th\'eorique, Universit\'e
de  Gen\`eve,}
\centerline{\em 24 Quai E. Ansermet, CH-1211  Gen\`eve}

\centerline{\em $^b$Center for Cosmology and Particle Physics,
Department of Physics,}
\centerline{\em New York University, New York,
NY, 10003, USA}

\centerline{}
\centerline{}

\vskip 1.cm

\begin{abstract}

We explore cosmology  in the decoupling limit of a
non-linear covariant extension  of Fierz-Pauli massive gravity
obtained recently in  arXiv:1007.0443.  In this limit the theory
is a scalar-tensor model of a unique form defined by symmetries.
We find that it admits a self-accelerated solution, with the
Hubble parameter set by  the graviton mass.
The negative pressure causing the acceleration is
due to a condensate of the helicity-0 component of the massive graviton,
and the  background evolution, in the approximation used, is
indistinguishable from the $\Lambda$CDM model.
Fluctuations about the self-accelerated   background are
stable for a certain range of parameters involved.
Most surprisingly, the fluctuation of the helicity-0 field above
its background decouples from an arbitrary source in the linearized theory.

We also show how massive gravity can  remarkably screen an arbitrarily large
cosmological constant in the decoupling limit, while evading issues with
ghosts. The obtained static solution is stable
against small perturbations, suggesting that the
degravitation of the vacuum energy is possible in the full theory.
Interestingly, however, this mechanism
postpones the Vainshtein effect to shorter distance scales. Hence,
fifth force measurements severely constrain the value of the
cosmological constant that can be neutralized,
making this scheme phenomenologically not viable
for solving the old cosmological constant problem. We
briefly speculate  on a possible way out of this issue.

\end{abstract}


\end{titlepage}

\newpage

\section{Introduction and summary}

The observed late-time acceleration of the Universe  \cite{acc},
and the Cosmological Constant problem (see reviews
\cite {Weinberg:1988cp,Polchinski}), remain two of the most
tantalizing, mutually connected puzzles
at the interface of particle physics and cosmology.

A promising approach to the late-time acceleration enigma
is to invoke  new degrees of freedom,  belonging  to the
gravitational field itself (as in massive gravity),
that give rise to the cosmic speed-up.
This framework postulates the existence of a new energy scale
-- set by the graviton mass -- which is very low;  nevertheless,
this scale is technically natural in the quantum-field theoretical sense.
This approach, as is known  by now, is challenging theoretically
(hence, is interesting),  and happens to have
robust observational predictions.

Such a scenario was first worked out  in a context of the
DGP model \cite{DGP} in Refs. \cite{Cedric,DDG}, where  the cosmic acceleration is due
to the  helicity-0 component of a five-dimensional graviton. Hence,
the solution is said to be self-accelerating.

Regretfully, in the context of DGP, the self-accelerating solution is
plagued by negative energy ghost-like states in the perturbative
approach \cite{Ratt,Koyama,Tanaka}, and despite the issue  of
whether or not the negative energy perturbations
could  be continued in the full nonlinear theory \cite{DGI}, the existence
of non-perturbative negative-energy solutions  \cite{Ratt,DW,GGcargese} makes
the self-accelerating branch  unsatisfactory (in spite of  the interesting
finding of Ref. \cite{Pujolas}
that the quasi-classical approach does not seem to reveal the
instabilities of this solution).

Certain generalizations of the DGP model, however, allow for
stable self-accele\-rating solutions, either by constructing an explicit braneworld
model \cite{deRham:2006pe} where the negative energy ghost disappears,
or by extending the decoupling limit of DGP to the ``galilean''
invariant interactions, \cite{nic}.

In this work, we show for the first time that a theory of massive gravity may produce a
self-accelerated geometry while being free of the problems that arise in the
self-accelerating branch of DGP. In particular, we will work in a certain
approximation in which the  helicity $\pm2$, $\pm 1$, and helicity-0 modes of
the massive graviton decouple from each other in the linearized theory, while
the nonlinear self-interactions,
and interactions between them,  are captured  by a few leading higher-dimensional terms
in the Lagrangian; this approximation constitutes the decoupling limit.

In this approximation,  we will show the existence of the self-accelerated solution,
around which small  fluctuations are stable. The acceleration is due to
a condensate of the
helicity-0  field,  which in the decoupling limit is reparametrization invariant.
On the other hand, since the helicity-0 is not an arbitrary scalar, but
descends from a full-fledged  tensor field, it has no potential,  but
enters the Lagrangian
via   very specific derivative terms fixed by symmetries \cite{CdRGG}. These terms
generate the negative pressure density which causes the accelerated expansion
with stable fluctuations, as will be discussed below.

From the observational point of view, the obtained self-accelerating
background is indistinguishable, in the approximation used,
from that of the $\Lambda$CDM model. As to the fluctuations, however, the helicity-0
could have introduced some differences.  For instance, at cosmological distance scales
it could have given an additional force leading to, {\it e.g.},  changes in the
growth of structure \cite{Roman,khoury}, while at shorter  scales still being
strongly screened via the Vainshtein mechanism \cite{Arkady}, guaranteeing the
recovery of General  Relativity with tiny departures
\cite{Arkady,DDGV},  which may also be measurable \cite{DGZ,Lue} in high-precision Laser
Ranging experiments \cite{Appolo} (for recent detailed studies of the Vainshtein mechanism
see Refs. \cite {Babichev}).  All the above takes place in the DGP model. However,
this is not what happens on the self-accelerated
background in the massive theory: Surprisingly-enough, the fluctuation of the
helicity-0 on this background decouples in the linearized approximation from
an arbitrary source!
Thus, the astrophysical sources need not excite this fluctuation, in which case
one recovers exactly the  $\Lambda$CDM  results.
It is likely, however,  that this similarity of the self-accelerated solution and its
fluctuations to the  $\Lambda$CDM results will not  hold beyond the decoupling limit
({\it i.e.} will not hold for the horizon-size scales).

\vspace{0.3cm}

Furthermore, if we wish to tackle the Cosmological Constant problem (CCP),
S. Weinberg's no-go theorem makes it impossible to find dynamical solutions
within General Relativity (GR) without involving fine-tuned parameters,
\cite{Weinberg:1988cp}.
The idea of infrared (IR) modification  of gravity, however addresses this puzzle by
accepting  a large vacuum energy  and modifying instead the gravitational
sector in the IR,  so that vacuum energy gravitates very weakly \cite {DGS}.
Such a source would not manifest itself as  strongly as naively anticipated in
GR, {\it i.e.} it would be {\it degravitated},  while all the  astrophysical sources
would exhibit the GR behavior \cite{ArkaniHamed:2002fu}. As shown in
Ref.~\cite{DGS,ArkaniHamed:2002fu}, one can think of degravitation as a promotion of
Newton's constant to a high pass filter operator thereby modifying the
effect of long wavelength sources such as a CC while recovering GR on shorter wavelengths.
In particular, theories of massive and resonance gravitons
exhibiting the high pass filter behavior to degravitate
the CC \cite{DGS}.  Moreover, it was shown in Ref.~\cite{Dvali:2007kt} that any causal
theory that can degravitate the CC is a theory of massive and resonance gravitons.

It is important to emphasize  that in theories of massive gravity
degravitation is a causal process (unlike more general theories
considered in \cite{ArkaniHamed:2002fu}). The real measure of whether or
not a source is degravitated is given
by its time evolution. 
During inflation for instance, the vacuum
energy driving the acceleration of the Universe will not be degravitated
for a long time. It is only after long enough periods of time that the IR modification of gravity 
kicks in and can effectively slow down an accelerated expansion \cite{DGS,ArkaniHamed:2002fu}.
Hence, a crucial ingredient for the
degravitation mechanism to work is the existence of a (nearly) static solution in
the presence of a cosmological constant towards which the geometry can relax at
late time (or after some long period of time). Indeed, Ref.~\cite{Dvali:2007kt}
studied linearized massive gravity  demonstrating  that in this approximation
degravitation takes place  after a long  enough period of time.

In this paper, we focus on the hard  mass case using the generalized Fierz-Pauli
theory of massive gravity, as derived in \cite{CdRGG}. 
We show that this model allows for static
solutions while evading any ghost issues at least in the decoupling limit.
In this framework an arbitrary vacuum energy can be neutralized by the
effective stress-tensor of the helicity-0 component of the massive graviton.
Small fluctuations around this solution  are shown below to be stable,
as long as this static solution exists.

Moreover, we find that the energy scale at which the  interactions of the
helicity-0 modes  become nonlinear is affected by the scale of the degravitated
cosmological constant --  the interaction scale being  higher for
larger values of the CC\footnote{In this work we will use interchangeably the notions of
vacuum energy and CC, although there could be a big difference between the two when it comes to
IR modified gravity \cite{DGS}.}.

On the one hand,  it is intriguing that the interactions of the helicity-0
can be kept  linear up to the energy scale  which is significantly higher
than what it  would have been in a theory without the CC. However, this very
same phenomenon also creates a problem by postponing Vainshtein's recovery of GR to
shorter and shorter distance scales.  As a result, the tests
of gravity impose a stringent upper bound on the vacuum energy that can be degravitated
in this framework without conflicting measurements of gravity.  Disappointingly,
this upper bound turns out to be of the order  the critical energy density of the
present-day Universe, $ (10^{-3} ~{\rm eV})^4$ -- the value that does not need to be
degravitated.

A possible way out of this difficulty may be to envisage a cosmological
scenario in which degravitation of the vacuum energy takes place before
the Universe enters the radiation dominated epoch -- say during
the inflationary period, or even earlier.  By the end of that epoch then the
cosmology should reset itself to continue evolution along the other branch
of the solutions that exhibits the standard early behavior followed by the
self-acceleration,   found  in the present work.
The existence of such a transition would depend on properties of the
degravitating solution in the full theory. Since we have no detailed
knowledge of this solution at the time of this writing,
we have no concrete mechanism to substantiate the above scenario.
Therefore, in what follows we will not rely on it.
Instead, we emphasize  that there still are  two
important virtues of the degravitating solution with the low
value of the degravitated CC:
(I) It is a concrete example of how degravitation could
work in four-dimensional theories of massive gravity without giving rise to ghost-like
instabilities.
(II) As we will show, the  degravitated solution with small values of CC
can be combined with the self-accelerated solution discussed above,
to give  a satisfactory solution that is in agreement with the existing
cosmological and astrophysical data.

Last but not least, the solutions found in the decoupling limit
do not necessarily imply the existence of the solutions with identical properties
in the full theory.  Nevertheless, the  decoupling limit solutions should capture
the local dynamics at scales well within  the present-day  Hubble four-volume,
as argued in \cite {nic}.  On the other hand, at  larger scales  the full solutions may
be very different from our ones.  These  differences would kick in at scales
comparable to the graviton Compton wavelength. Therefore, our solutions should
manifest themselves at least as transients lasting long cosmological times.

Organization of the paper is as follows. In section~\ref{Formalism}, we review
the generalized Fierz-Pauli theory of massive gravity and discuss its ghostless
decoupling limit. We then start by focusing on self-accelerating solutions
in section~\ref{SelfAcc}, first deriving the background solutions, then
testing their stability, and finally studying the implications for late-time
cosmology. We then explore the cosmology in
the presence of a cosmological constant in section~\ref{Degravitation}, proving the
existence of a stable degravitating branch of solutions,  and analyzing the stability
of the de Sitter branch. Brief discussions of the degravitating solution
are given at the end of section~\ref{Degravitation}.

\section{The Formalism}
\label{Formalism}

Search for a consistent theory of a massive spin-2 field goes back to the
original work of Fierz and Pauli \cite{FP}.  Whereas any massive gravity should reduce
to the Fierz-Pauli (FP) theory at the quadratic level \cite{Nieu},
a generic nonlinear extension exhibits the sixth
degree of freedom -- the so-called Boulware-Deser  (BD) ghost \cite {BD}.
This sixth mode produces  severe
instabilities on cosmological backgrounds \cite {GGruzinov}, as well as
on locally nontrivial asymptotically flat backgrounds
(such as that of a point source, for instance) \cite {AGS,Creminelli1,DeffayetRombouts}.

This problem is usually  related  to the helicity-0 sector of massive
theories \cite {AGS}. The latter can efficiently be studied in the decoupling
limit, where the sixth mode is hidden in higher-derivative
nonlinear terms for the helicity-0 \cite {AGS,Creminelli1,DeffayetRombouts}.
Such terms make the Cauchy problem ill-defined,  unless additional initial
data are supplied. This corresponds to an additional, sixth, degree of freedom which
shows up as a ghost-like linear mode on various backgrounds mentioned above.

Up until recently it was thought that the cancellation of the higher-derivative
nonlinear terms for the helicity-0 was not possible \cite {Creminelli1}.
However, recently an explicit
construction was given in Ref. \cite {CdRGG} in which all the nonlinear terms
for the helicity-0 with more that two time derivatives cancel.
Below we briefly review these results  and recast them in a more convenient form.
We refer to Ref. \cite{CdRGG} for  more detailed discussions.

Consider a 4D covariant theory of a spin-2 field \cite {AGS},  which, once expanded
on Minkowski space-time gives a graviton of mass $m$:
\ba
{\cal L} =  M^2_{\rm Pl} \sqrt{-g}  R -   \frac{ M^2_{\rm Pl} m^2}{4}
\sqrt{-g} \left ({U}_2(g,H)+{U}_3(g,H)+{U}_4(g,H)+ {U}_5(g,H)\cdots\right) \,.
\label{PF}
\ea
Here $U_i$'s denote  the mass and potential terms  of
$i^{\rm th}$ order in $H\mn$
\ba
\label{PFS}
{U}_2(g,H)&=&H^2_{\mu\nu}-H^2\,,\\
{U}_3(g,H)&=&c_1 H\mn^3+c_2 H H\mn^2+c_3 H^3\label{L3}\,,\\
{U}_4(g,H)&=&d_1 H\mn^4+d_2 H H\mn^3+d_3 H\mn^2H_{\alpha\beta}^2+
d_4 H^2 H\mn^2+d_5 H^4\label{L4}\,,\\
{U}_5(g,H)&=&f_1 H\mn^5+f_2 H H\mn^4+f_3 H^2 H\mn^3+
f_4 H_{\alpha\beta}^2 H\mn^3 \nonumber \\
&+&f_5H (H\mn^2)^2+f_6H^3 H\mn^2+f_7 H^5 \label{L5}\,.
\ea
Index contractions are performed using the inverse metric $g^{\mu\nu}$; the coefficients
$c_i,d_i$ and $f_i$ are {\it a priori} arbitrary. The tensor $H\mn$ is not an independent entity;
it is related to the metric tensor as  $ H\mn =g\mn -
\eta_{ab}\partial_\mu \varphi^a \partial_\nu \varphi^b\,,$
where $a,b =0,1,2,3,$  $\eta_{ab}={\rm diag} (-1,1,1,1)$, and $H\mn$ is a
covariant tensor as long as the four fields  $\varphi^a$ transform  as scalars  under  a
change of coordinates \cite {AGS}. Hence, the potential terms
in (\ref {PF})   can be rewritten as functions  of
the metric $g$ and  the specific combination of the
four scalars $\varphi^a$, as $U(g,\Sigma)$, where $\Sigma_{\mu\nu}=\eta_{ab}\partial_\mu
\varphi^a \partial_\nu \varphi^b $. However, we will not be exploiting
the latter representation in the present work.  Instead,  following \cite {AGS}
we expand  $\varphi^a$ in terms of the  coordinates $x^\alpha$, and the field $ \pi^\alpha $, as
$ \varphi^a= (x^\alpha-\pi^\alpha)\, \delta^a_\alpha \,,$ and using
the convention,  $g_{\mu\nu} = \eta_{\mu\nu} + {h_{\mu\nu}}/{\mpl} $, we obtain
\ba
\label{Hmn}
H\mn=\frac{h\mn}{\mpl}+\partial_\mu \pi_\nu + \partial_\nu \pi_\mu -
\eta_{\alpha\beta}\partial_\mu \pi^\alpha \partial_\nu \pi^\beta\,.
\ea
The $\pi_\alpha $'s  represent the St\"uckelberg
fields that transform under reparametrization to guarantee that the tensor
$H$ in (\ref {Hmn}) transforms covariantly. In the unitary gauge one could put
$\pi_\alpha =0$ (or, $\varphi^a= x^\alpha\delta^a_\alpha$), in which case (\ref{PF}) reduces
to the standard FP theory extended by a potential for the field $h_{\mu\nu}$.
However, this is not a convenient way of dealing with these
degrees of freedom. Instead, it is more instructive to retain $\pi_\alpha$
and fix a gauge for $h_{\mu\nu}$.

The theory (\ref {PF}) was studied in detail in \cite{CdRGG,deRham:2010gu}, and
a two-parameter family  of the coefficients was
identified for which no sixth (ghost) degree of freedom  arises
in the  decoupling limit\footnote
{Interestingly, a recently proposed extension of General Relativity by an
extra auxiliary dimension \cite{GG, CdR},  automatically generates the coefficients
from this family at least up to  the cubic order.}.  In these theories the
higher derivative nonlinear terms either cancel out, or organize
themselves into total derivatives. For these ghostless  theories, the decoupling
limit is defined as follows\footnote{By ``ghostless" we mean a theory
with no ghost at least in the decoupling limit, implying  that even
if the BD ghost exists in the full theory, it must
have a mass larger than the scale $\Lambda_3$, \cite{CdRGG}.}
\beq
m \to 0, ~~~\mpl \to \infty, ~~~  \Lambda_3 = (\mpl m^2)^{1/3} ~~{\rm fixed}.
\label{declim3}
\eeq
In what follows, we will focus on the helicity-2 and helicity-0 modes, and
ignore the helicity-1 modes as they do not couple to a conserved stress-tensor at the linearized level,
and, therefore, can be set to zero self-consistently (see, however,
important comments on this at the end of section 3.2).

We therefore use the following decomposition for $H_{\mu\nu}$
in terms of the canonically normalized helicity-2
and helicity-0 fields after setting $\pi_a=\partial_a \pi / \Lambda_3^3$
\ba
H\mn&=&\frac{h\mn}{\mpl}+\frac{2 \partial_\mu \partial_\nu\pi  }{\Lambda_3^3}
-\frac{\partial_\mu\partial^\alpha\pi\partial_\nu\partial_\alpha \pi  }
{\Lambda_3^6}.
\ea
Then, one can show by direct calculations \cite {CdRGG}
that the Lagrangian (\ref{PF}) reduces in the decoupling limit to
the following expression
\beq
\mathcal{L}=-\frac{1}{2}
h^{\mu\nu}\mathcal{E}^{\alpha\beta}_{\mu\nu} h_{\alpha\beta}+
 h^{\mu\nu}\sum_{n=1}^3 \frac{a_{n}}{\Lambda_3^{3(n-1)}} X^{(n)}_{\mu\nu}[\Pi],
\label{lagr}
\eeq
where the first term represents the usual kinetic term for the
graviton, $a_1=-1/2$,  and $a_{2,3}$ are two arbitrary
constants, related to the two parameters from the set $\{c_i,d_i\}$ which characterize
a given ghostless theory of massive gravity.
The expression $\(\mathcal{E}h\)\mn$  denotes the linearized Einstein
operator acting on  $h_{\mu\nu}$ defined in the standard way:
$\mathcal{E}^{\alpha\beta}_{\mu\nu} h_{\alpha\beta}=-\frac12 (\Box h\mn-
\p_\mu\p_\alpha h^\alpha_{\, \nu}-\p_\nu\p_\alpha h^\alpha_{\, \mu}+\p_\mu\p_\nu h
-\eta\mn \Box h + \eta\mn \p_\alpha \p_\beta h^{\alpha\beta} )$.

The three symmetric tensors  $X^{(n)}_{\mu\nu}[\Pi]$ are composed of the second
derivative of the helicity-0 field $\Pi_{\mu\nu}\equiv \partial_\mu \partial_\nu \pi$.
The $\pi$ field ends up  being gauge invariant in the decoupling limit.
In order to maintain reparametrization invariance of the full Lagrangian
the tensors  $X^{(n)}_{\mu\nu}[\Pi]$  should be identically conserved.
These properties uniquely determine the expressions for $X^{(n)}_{\mu\nu}$ at
each order of non-linearity. The obtained expressions  agree  with the results of
the direct  calculations of Ref.  \cite {CdRGG}. A convenient parametrization
for the tensors $X^{(n)}_{\mu\nu}$ which we adopt in this work is as follows:
\begin{align}
X^{(1)}_{\mu\nu}[\Pi]={\varepsilon_{\mu}}^{\alpha\rho\sigma}
{{\varepsilon_\nu}^{\beta}}_{\rho\sigma}\Pi_{\alpha\beta}, \quad  \nonumber \\
X^{(2)}_{\mu\nu}[\Pi]={\varepsilon_{\mu}}^{\alpha\rho\gamma}
{{\varepsilon_\nu}^{\beta\sigma}}_{\gamma}\Pi_{\alpha\beta}
\Pi_{\rho\sigma}, \nonumber \\
X^{(3)}_{\mu\nu}[\Pi]={\varepsilon_{\mu}}^{\alpha\rho\gamma}
{{\varepsilon_\nu}^{\beta\sigma\delta}}\Pi_{\alpha\beta}
\Pi_{\rho\sigma}\Pi_{\gamma\delta}\,.
\label{Xs}
\end{align}
The remarkable property of (\ref{lagr}) is that it represents the
\textit{exact} Lagrangian (excluding the helicity-1 part) in the
decoupling limit:  All the higher than quartic terms vanish
in this limit, making (\ref{lagr}) a unique theory to which
any nonlinear, ghostless extension of massive gravity should
reduce in the decoupling limit \cite {CdRGG}.

If external sources are introduced, their stress-tensors then couple to the
physical metric $h_{\mu\nu}$.  In the basis used in (\ref {lagr})
there is no direct coupling of $\pi$ to the stress-tensors.
Hence, the Lagrangian (\ref {lagr})  is invariant
w.r.t. the shifts,  and  the ``galilean'' transformations in the
internal space of the $\pi$ field,
$\partial_\mu \pi\to \partial_\mu \pi + v_\mu$,
where $v_\mu$ is a constant four-vector.
The  latter invariance guarantees that there is no mass
nor  potential terms generated for $\pi$ by the loop corrections.

The tree-level coupling  of $\pi$ to
the sources arises only after diagonalization:
The  quadratic mixing $h^{\mu\nu}X^{(1)}_{\mu\nu}$, and
the cubic interaction $h^{\mu\nu}X^{(2)}_{\mu\nu}$,
can be diagonalized by a nonlinear transformation of $h_{\mu\nu}$,
that generates  the following coupling of  $\pi$ \cite{CdRGG}
\beq
{ 1\over \mpl} \left ( -2 a_1\eta_{\mu\nu} \pi +  {2a_2\partial_\mu \pi \partial_\nu
\pi \over  \Lambda_3^3}
\right ) T^{\mn}\,.
\label{invcoupling}
\eeq
Moreover, the above transformation also generates all the Galileon
terms for the helicity-0 field,  introduced in a different context
in Ref. \cite{nic}\footnote{The usual coupling
of the Galileon field to the stress-tensor, $\pi T$ ,
considered in the generic  Galileon theories, violates
the  ``galilean'' invariance that is needed to protect
the Galileon field from acquiring a mass and potential terms.
Here however, the invariance is manifest before diagonalization
and the theory is protected.}.

Since the Galileon terms are known to exhibit the
Vainshtein recovery of GR at least for static sources \cite{nic},
so does the above theory with  $a_3=0$.
The quartic interaction  $h^{\mu\nu}X^{(3)}_{\mu\nu}$,  however, cannot be
absorbed by any local redefinition of $h_{\mu\nu}$.  It is still
expected though to admit the Vainshtein mechanism.

However, as we will show in the next section,
on the self-accelerated background  the fluctuation of the helicity-0
field decouples from an arbitrary source, making the predictions
of the theory consistent with GR already in the linearized approximation.
This decoupling is a direct consequence of the self-accelerated
background and the specific form of the coupling
(\ref {invcoupling}).

\section{The Self-Accelerated Solution}
\label{SelfAcc}

The universality of the decoupling limit Lagrangian
(\ref {lagr}) for the class of ghostless massive gravities,
suggests the possibility  of a fairly model-independent phenomenology of
the massive theories that should be captured by the
limiting Lagrangian (\ref {lagr}). In the present section, we
will be interested
in the cosmological solutions  in these theories. We will directly work in the
decoupling limit, which implies scales much smaller than the Compton wavelength
of the graviton. In the case of the self-accelerated de Sitter solution
for instance,
this corresponds to probing physics within the Hubble scale, which as one would
expect, is set  by the value of the graviton mass.

\subsection{The solution in the decoupling limit}

Below we look for homogeneous and isotropic solutions
of the equations of motion that follow from the Lagrangian
(\ref {lagr}). The helicity-0 equation of motion reads as follows:
\beq
\partial_\alpha\partial_\beta h^{\mu\nu} \left (a_1{\varepsilon_{\mu}}^{\alpha\rho\sigma}{{\varepsilon_\nu}^{\beta}}_{\rho\sigma}+2\frac{a_2}{\Lambda_3^3}{\varepsilon_{\mu}}^{\alpha\rho\sigma}{{\varepsilon_\nu}^{\beta\gamma}}_{\sigma}\Pi_{\rho\gamma}+3\frac{a_3}{\Lambda_3^6}{\varepsilon_{\mu}}^{\alpha\rho\sigma}{{\varepsilon_\nu}^{\beta\gamma\delta}}
\Pi_{\rho\gamma}\Pi_{\sigma\delta}\right )=0\,,
\label{1}
\eeq
while variation of the Lagrangian w.r.t.  the helicity-2 field gives
\beq
- \mathcal{E}^{\alpha\beta}_{\mu\nu}
h_{\alpha\beta}+\sum_{n=1}^3 \frac{a_{n}}{\Lambda_3^{3(n-1)}} X^{(n)}_{\mu\nu}[\Pi]=0.
\label{2}
\eeq
We are primarily interested in the self-accelerated solutions of the
system (\ref{1})-(\ref{2}). For an observer at the origin of the coordinate system,
the de Sitter metric can locally
({\it i.e.},  for times $t$,  and physical distances $|\vec{x}|$, much smaller than
the Hubble scale $H^{-1}$) be written as a small perturbation over
Minkowski space-time \cite{nic}
\beq
\d s^2=[1-\frac{1}{2} H^2x^{\alpha}x_{\alpha}]\eta_{\mu\nu}\d x^\mu \d x^\nu.
\label{ds}
\eeq
The linearized Einstein tensor for the (dimensionless)
metric (\ref{ds}) is given by
\beq
G^{\rm lin}_{\mu\nu}={1\over \mpl} \mathcal{E}^{\alpha\beta}_{\mu\nu}
h_{\alpha\beta}=-3 H^2 \eta_{\mu\nu}.
\eeq
For the helicity-0 field we look for the solution of the following
isotropic form
\beq
\label{pi_ansatz}
\pi=\frac{1}{2}\, \kk \Lambda_3^3 x^\alpha x_\alpha+b \Lambda_3^2 t+c \Lambda_3\,,
\eeq
where $\kk, b$ and $c$ are three dimensionless constants.

The equations of motion for the helicity-0 and helicity-2 fields (\ref{1})-(\ref{2}),
therefore,  can be recast in the following form
\beq
H^2 \left (  -\frac{1}{2}+2a_2 \kk+3 a_3 \kk^2 \right )=0,~~~~~~\label{3}\\
 \mpl H^2=2 \kk \Lambda_3^3  \left[ -\frac{1}{2}+   a_2 \kk+a_3\kk^2 \right]~.
 \label{3'}
\eeq
Solving the quadratic equation (\ref{3}) for $\kk$ (for $H\neq 0$), we obtain the Hubble
constant of the self-accelerated solution  from (\ref{3'}). Its magnitude,
$H^2\sim\Lambda^3_3/\mpl=m^2$,  is set by the graviton mass, as expected
(positivity of $H^2$ is one of the conditions that we will
be demanding  below).  It is not hard to convince oneself that there exists a whole set of
self-accelerated solutions, parametrized by $a_2$ and $a_3$. This range, however,
will be restricted further by the requirement of stability of the solution, which is the
focus of the next section.

Before doing so, let us briefly analyze the four scalars $\varphi^a$. Using the ansatz, \eqref{pi_ansatz},
their expression is given by
\ba
\varphi^a=(1-q) x^\alpha \delta^a_\alpha\,,
\ea
if we set $b=0$. Thus the four scalars vanish in the special case of $q=1$,
and the metric  is $g\mn=H\mn$, so that the Lagrangian considered in \eqref{lagr}
reduces to standard GR plus a CC (at least at the background level).
This happens only if the parameters of our theory are such that
$a_3=\frac{1}{6}-\frac23 a_2$, which is not the regime we will be interested in --
we will indeed show in what follow that the stability of the self-accelerated background
implies  $q\ne 1$.

\subsection{Small perturbations and stability}

Here we  investigate the constraints that the requirement of stability
imposes on a possible background. Let us adopt a particular
solution of the system (\ref{3})-(\ref{3'}) and consider perturbations on
the corresponding de Sitter background
 \beq
 h_{\mu\nu}= {h}^{b}_{\mu\nu}+\chi_{\mu\nu}, \quad \pi=\pi^{b}+\phi,
 \eeq
where the superscript $b$ denotes the corresponding background values.
The Lagrangian for the perturbations (up to a total derivative) reads as follows
\beq
\mathcal{L}&=& -\frac{1}{2}\chi^{\mu\nu}\mathcal{E}^{\alpha\beta}_{\mu\nu}
\chi_{\alpha\beta}+6(a_2+3a_3\kk)\frac{H^2 \mpl}{\Lambda_3^3}\phi
\Box\phi-3a_3\frac{H^2 \mpl}{\Lambda_3^6}(\partial_\mu\phi)^2\Box\phi \nonumber \\
&+&\frac{a_2+3 a_3 \kk}{\Lambda_3^3}\chi^{\mu\nu} X^{(2)}_{\mu\nu}[\Phi]
+\frac{a_3}{\Lambda_3^6}\chi^{\mu\nu} X^{(3)}_{\mu\nu}[\Phi]+\frac{\chi^{\mu\nu}T\mn}{\mpl}\,,
\label{4}
\eeq
where $\Phi$ denotes the four-by-four matrix with the elements
$\Phi_{\mu\nu} \equiv \partial_\mu \partial_\nu \phi$.
The first term in the first line of the above expression  is the Einstein  term
for  $\chi_{\mu\nu}$, the  second term is a
kinetic term for the scalar,  and the third one is the cubic Galileon.
The second line contains cubic and quartic interactions between
$\chi_{\mu\nu}$ and $\phi$, which are identical in form  to the corresponding
terms in the decoupling limit on Minkowski space-time (\ref{lagr}).
None of these interactions therefore lead to ghost-like instabilities
\cite{CdRGG}, as long as the $\phi$ kinetic term is positive definite.

Most interestingly, however, there is no quadratic mixing term between
$\chi$ and $\phi$ in (\ref {4}). Since it is only $\chi_{\mu\nu}$
that couples  to external sources $T\mn$ in the quadratic approximation, then
there will not be a quadratic coupling of $\phi$  to the sources  generated in
the absence of the
quadratic $\chi -\phi$ mixing.  Therefore, for arbitrary external sources, there exist
consistent solutions for which the fluctuation of the helicity-0 is not excited, $\phi=0$.
On these solutions one exactly recovers  the results of the linearized GR. The
above phenomenon
provides a mechanism of decoupling  the helicity-0 mode from  arbitrary external sources!
This mechanism is a universal property of the self-accelerating solution in ghostless
massive gravity.

Hence, there are no instabilities in (\ref{4}),  as long as $a_2+3a_3\kk>0$.
The latter condition, along with the requirement of positivity of $H^2$,
and the equations of motion (\ref{3}),  requires that the following system
be satisfied:
\beq
-\frac{1}{2}+2a_2 \kk+3 a_3 \kk^2=0, ~~~~~~~~~~~~~~~~~~\nonumber\\
 \mpl H^2=2 \kk \Lambda_3^3\left[a_2 \kk+a_3\kk^2-\frac{1}{2}\right]>0,~~~
 a_2+3a_3\kk>0, \nonumber
\eeq
for the self-accelerating solution to be physically meaningful.
The above  system can be solved. The solution is given as follows
\beq
a_2<0,~~-\frac{2a_2^2}{3}<a_3<-\frac{a_2^2}{2},
\label{bounds}
\eeq
while the Hubble constant and  $\kk$ are given by the following expressions
\beq
H^2= m^2[2 a_2 \kk^2+2 a_3\kk^3-\kk]>0,~~\kk=-\frac{a_2}{3a_3}+
\frac{(2a_2^2+3a_3)^{1/2}}{3\sqrt{2}a_3}.
\eeq
It is clear from (\ref{bounds}), that the undiagonalizable interaction
$h^{\mu\nu}X^{(3)}_{\mu\nu}$ plays a crucial role for the stability of this
class of solutions: All theories without this term ({\it i.e.} the ones with
$a_3=0$) would have ghost-like instabilities on the self-accelerated background.
Notice as well that in the regime \eqref{bounds} none of the scalars $\varphi^a$ vanish,
and our model therefore differs from GR with a CC.

We therefore conclude that there exists a well-defined  class of massive theories
with the parameters satisfying the conditions (\ref{bounds}),  which
propagate no ghosts on asymptotically flat backgrounds, and
also admit stable self-accelerated solutions in the decoupling limit.

As we mentioned before, the helicity-1 field enters only quadratically,
or in higher order terms in the Lagrangian, and hence, can consistently be set
to be zero ({\it i.e.} it does not need to be excited by any other fields).
Nevertheless, once a background configuration  for the helicity-0 field
is switched on,  the higher-dimensional mixed terms of  the helicity-0 and
helicity-1  could in principle flip the sign of the Maxwell kinetic term,
giving rise to a vector ghost that would  only enter the Lagrangian quadratically
or in higher powers; this field  would couple  to other fields at the nonlinear level.
This certainly would not be a satisfactory state of affairs.

By restoring back the helicity-1 field in our expressions, and performing direct
calculations we have shown that in  the $n^{\rm th}$ order in nonlinearities, where $n\le 6$,
the coefficient of the Maxwell term on the self-accelerated background  is proportional
to $(-{1\over 2} + 2a_2 q + 3 a_3q^2)$, up the corrections that are of
the $(n+1)^{\rm th}$ order, \cite{deRham:2010kj}. Hence, up to these corrections,  the Maxwell term
vanishes on the self-accelerated background!

If this were the full story we would get a theory of helicity-1 coupled infinitely strongly
to the fluctuation of helicity-0 in the decoupling limit.
However, quantum loop corrections will  necessarily generate
a nonzero Maxwell term,   as it is not protected  by any symmetries.  In these loops propagate the
helicity-0 mode,  as well as  the matter fields to which the helicity-1 couples nonlinearly
(for instance, one of the couplings being,  $\partial_\mu A_\alpha \partial_\nu A_\alpha T^{\mu\nu}$).

Then, interpreting the value of the tree-level coefficient of the Maxwell term (which is zero) as
an infinite value of the inverse of the running $U(1)$ coupling at some UV scale $\Lambda_{UV}\ge \Lambda_3$,
we obtain  that at lower scales the coupling constant has a positive value
as long as the theory is not asymptotically free (in other words,
the $U(1)$ coupling would have a Landau pole at some high scale $\Lambda_{UV}$)\footnote{Alternatively,
if the particle content is such that the theory has a negative beta function, then
the infinite value of the coupling constant should be attributed to some far IR scale,
$\Lambda_{IR} \ll \Lambda_3$, and at any scale greater than  $\Lambda_{IR} $ the helicity-1 theory would
have a finite positive coupling square.}.
Hence, the helicity-1 sector would not have a ghost, but the scale at which
it would become nonlinearly  interacting (the Vainshtein scale) would be parametrically (logarithmically)
smaller than  $\Lambda_3$. Since the helicity-1  field  does not have to be excited by
any source, this will not be a concern for us.

\subsection{Late-time cosmology}

In this subsection we discuss  the relevance of the results obtained above for the
late-time local cosmological evolution of the Universe. As seen from the decoupling limit
Lagrangian (\ref{lagr}), the helicity-0 mode $\pi$
provides an effective stress-tensor  that is ``felt'' by
the helicity-2 field:
\beq
T^\pi_{\mu\nu}=\mpl\sum_{n=1}^3 \frac{a_{n}}{\Lambda_3^{3(n-1)}} X^{(n)}_{\mu\nu}[\Pi]
=-6 \kk \mpl\Lambda_3^3  \left[-\frac{1}{2}+   a_2 \kk+a_3\kk^2 \right]
\eta_{\mu\nu}\,.
\label{ve}
\eeq
It is this stress-tensor that provides the negative pressure density required
to drive the acceleration of the Universe. Supplemented by the matter density
contribution, it leads to the usual $\Lambda$CDM - like cosmological  expansion
of the background in the sub-horizon approximation used here. This is clear form the fact
that the stress-tensor (\ref {ve}) gives rise to a de Sitter background as was shown in the
previous subsection. Hence, in the comoving coordinate system -- which differs from the one
used above -- the invariant de Sitter space will be the self-accelerating solution.

All this can be reiterated by performing an explicit coordinate transformation
to the comoving coordinates. This will be done in two steps.
In the so-called Fermi normal coordinates, the FRW metric can be
locally written in space and \textit{for all times},
as a small perturbation over Minkowski space-time:
\beq
\d s^2=-[1-(\dot H+H^2)\bold{x}^2]\d t^2+
\left[1-\frac{1}{2}H^2 \bold{x}^2\right]\d\bold{x}^2=
\left(\eta_{\mu\nu}+h^{\rm FRW}_{\mu\nu}\right)\d x^\mu \d x^\nu
\label{FRW},
\eeq
where the corrections to the above expression are suppressed by higher powers of
$H^2\bold{x}^2$. The Fermi normal coordinates, on the other hand,
are related to those  used in (\ref{ds}) (in which the FRW metric is a small conformal deformation of
Minkowski space-time), by an infinitesimal gauge transformation \cite{nic}. The latter does not change
the expression (\ref {ve}), since  $T^\pi_{\mu\nu}$ is invariant under infinitesimal
gauge transformations in the decoupling limit. On the other hand, the Fermi normal
coordinates can be transformed  into the standard comoving
coordinates $(t_c,\bold{x}_c)$  as follows  \cite{nic}
\beq
t_c=t-\frac{1}{2}H(t)\bold{x}^2,~~~~
\bold{x}_c=\frac{\bold{x}}{a(t)}\left[1+\frac{1}{4}H^2(t)\bold{x}^2\right].
\eeq
The stress-tensor of a perfect fluid,  $T_{\mu\nu}=\text{diag}(\rho(t_c),
a^2(t_c)p(t_c)\delta_{ij})$, transforms under this change of coordinates
(at the leading order in $H^2 \bold{x}^2$) into the following expression
\[
 T_{\mu\nu} =
 \begin{pmatrix}
  \rho & -H(\rho+p)x^i \\
  -H(\rho+p)x^i & p\delta_{ij}    \\
  \end{pmatrix},
\]
where all quantities in the latter expression are evaluated at time $t$.
Note that the off-diagonal entries of the stress-tensor for the cosmological
constant vanish in the  Fermi normal coordinates, the same is
true for $T^\pi_{\mu\nu}$ as well. Hence, in all coordinate systems used
the expressions for the stress-tensor on the self-accelerated solution
is given by  (\ref {ve}).

Not surprisingly, the corresponding cosmological equations
coincide with the conventional ones for the $\Lambda$CDM model,
with the cosmological constant set by the mass of the graviton
\beq
H^2=\frac{\rho}{3\mpl^2}+\frac{C^2 m^2}{3}, ~~~~~~~~~~~~~~~~~\\
\dot H+H^2=\frac{\ddot a}{a}=-\frac{1}{6\mpl^2}(\rho+3p)+\frac{C^2 m^2}{3}.
\eeq
Here $\rho$ and $p$ denote the energy and pressure densities
of matter and/or radiation, and $C^2\equiv 6 \kk
\left[ -\frac{1}{2}+  a_2 \kk+a_3\kk^2 \right] $ is a constant
that appears in (\ref {ve}).

As already mentioned, irrespective of the
completion (beyond the Hubble scale) of the self-accelerated solution,
it is locally indistinguishable from the $\Lambda$CDM model.
At the horizon scales, however, it is likely that these two
scenarios will depart from each other: As emphasized in  the first section, the
solutions found in the decoupling limit do not necessarily imply the existence of full solutions
with identical properties.   Moreover, the decoupling limit Lagrangian
(\ref {lagr}) is derived from the full theory by dropping
certain total derivative terms (see \cite {CdRGG}),  implying
solutions that decay fast-enough  at infinity. On the
other hand, the solutions that we found in this section are given in
the coordinate system where the fields grow  at large distance/time scales.
If these solutions are to be continued into the full theory,
the latter should have an appropriate large scale behavior in this
coordinate system.  A given solution in the decoupling limit can just
be a transient state of the full solution. Significant deviations of the latter  from
the former  should kick in at distance/time scales comparable to the graviton
Compton wavelength.

\section{Screening the Cosmological Constant}
\label{Degravitation}

\subsection{Degravitation in generic theories of massive gravity}

One explicit realization of degravitation is expected to occur in
massive gravity, where gravity is weaker in the IR, and the
graviton mass could play the role of a high-pass filter \cite{DGS}.
In the approach of \cite{DGS}, the original theory was  formulated as
a higher dimensional  model, the 4D reduction of which can be though  as
massive/resonance gravity with the mass term promoted into a specific differential
operators determined by the underlying higher-dimensional construction.

A  more general approach was  adopted in Ref. \cite{Dvali:2007kt},
where  the graviton mass was also promoted to an operator
parameterized  by  a continuous parameter $\alpha$
\ba
m^2(\Box)=m_0^{2(1-\alpha)}\Box^{\alpha}\,,
\ea
the inverse graviton propagator is typically of the form
\ba
\mathcal G^{-1} \sim \Box-m_0^{2(1-\alpha)}\Box^{\alpha}\,,
\ea
so that for $\alpha<1$, gravity is weaker beyond wavelengths comparable to the
graviton Compton wavelength $m_0^{-1}$.

To take this mechanism a step further, the reliability of this
argument within the non-linear regime is hence crucial.
A trick to manifest the key interactions that arise in massive gravity is to work
in the decoupling limit, where the usual GR interactions are suppressed, while the
interactions of the new degrees of freedom are emphasized.
This approach was first derived in Ref.~\cite{Dvali:2007kt},
which we discuss first before turning to our considerations.
As mentioned above, this limit is obtained by taking $\mpl \to \infty$ and $m\to 0$.
However unlike in the decoupling limit of the theory  discussed in the previous section, the nonlinear
dynamics in a generic model of massive gravity is governed by the scale
 \ba
\Lambda_\star^{5-4\alpha}=\mpl m^{4(1-\alpha)}\,.
\ea
In such models, it has been shown \cite{Dvali:2007kt} that the helicity-0 ($\pi$)
and -2 ($\bar h\mn$) modes satisfy the following equations in the decoupling limit,
\ba
&& -\mathcal{E}^{\alpha\beta}\mn\bar h_{\alpha\beta}=-\frac1{\mpl} T\mn \,,\\
\label{eqPi_old}
&& 3 \Box \pi-\frac{18}{\Lambda_\star^{5-4\alpha}}
\(3\Box(\Box^{1-\alpha}\pi)^2+\cdots\)=-\frac{T}{\mpl}\,,
\ea
where the physical metric is given by $g\mn=\eta\mn+(\bar h\mn+\pi \eta\mn)/\mpl$.
In the presence of a cosmological constant, $T\mn=-\lambda \eta\mn$,
the solution for the helicity-2 mode is
\ba
\bar h\mn=-\frac{\lambda}{6\mpl} x_\beta x^\beta \, \eta\mn\,,
\ea
which is the usual GR solution. One can now check the condition for the existence of
a (nearly) static solution towards which the geometry can relax at late times. In the language
of the decoupling limit, this would happen if the helicity-0 mode compensates the helicity-2 mode
contribution  $\pi \eta\mn=-\bar h\mn$ to maintain the geometry flat $g\mn=\eta\mn$.
However the configuration $\pi=\lambda x^2 /6\mpl$ is precisely the solution
of \eqref{eqPi_old} when the higher interactions vanish,
\ie $6\mpl \Box \pi=-T=8\lambda$. As shown  in \cite{Dvali:2007kt},
such interactions cancel for $\pi\sim x^2$ only if $\alpha<1/2$, hence
implying that a generic theory of massive gravity amended with a nonzero
CC can only have a static solution when $\alpha<1/2$.
In particular, in this language the DGP model \cite{DGP} corresponds to $\alpha=1/2$ (see Ref.~\cite{Ratt}, but also \cite{GGIglesias})
hence explaining why this model does not bear static solutions with a brane tension, while promoting it to higher dimensions corresponds to a theory
with $\alpha \to 0$ for which the usual codimension-two conical solutions
can accommodate a tension  without acceleration,
\cite{Gabadadze:2003ck,deRham:2007xp,deRham:2009wb,Padilla:2010de,Padilla:2010tj}.

The above results hold true for a generic theory of massive gravity.
We now focus the analysis of the ghostless theory \cite{CdRGG} reviewed
in section~\ref{Formalism}, which strictly speaking are not captured by the
above $\alpha$ parametrization.  The key difference in the ghostless case is
that interactions for the helicity-0 mode are governed by the larger coupling
scale $\Lambda_3>\Lambda_\star$. The form of these interactions in the ghostless
theory,  as well as the specific couplings to matter,  play a crucial role in
accommodating a degravitating branch of solutions, and this without being
plagued by any instability at least in the decoupling limit.

\subsection{Degravitation in ghostless massive gravity}

For convenience we start by recalling the decoupling limit
Lagrangian of (\ref{lagr}) coupled to an external source
\ba
\mathcal{L}=-\frac{1}{2}h^{\mu\nu}\mathcal{E}\mn^{\alpha\beta}h_{\alpha\beta}
+ h^{\mu\nu}\sum_{n=1}^3 \frac{a_{n}}{\Lambda_3^{3(n-1)}} X^{(n)}_{\mu\nu}[\Pi]
+\frac1{\mpl} h^{\mu\nu} T\mn\,.
\ea
The equations of motion for the helicity-0 and 2 modes are then
\ba
\label{eqh}
-\mathcal{E}\mn^{\alpha\beta}h_{\alpha\beta}+\sum_{n=1}^3 \frac{a_{n}}
{\Lambda_3^{3(n-1)}} X^{(n)}_{\mu\nu}[\Pi]=-\frac1{\mpl} T\mn \,,
\ea
and
\ba
\label{eqpi}
\big(a_1+\frac{a_2}{\Lambda_3^3}\Box \pi+\frac{3a_3}
{2\Lambda_3^6}\([\Pi]^2-[\Pi^2]\)\big)\,
\Big[\Box h-\p_\alpha\p_\beta h^{\alpha\beta}\Big]\nn\\
 +\frac{1}{\Lambda_3^3}\big(a_2\Pi\mn-3 \frac{a_3}
{\Lambda_3^3}\(\Pi^2\mn-\Box \pi \Pi\mn\)\big)\Big[2\p^\mu\p_\alpha h^{\alpha\nu}
-\Box h^{\mu\nu}-\p^\mu\p^\nu h\Big]\nn\\
-\frac{3a_3}{\Lambda^6_3}\(\Pi_{\mu\alpha}
\Pi_{\nu\beta}-\Pi_{\mu\nu}\Pi_{\alpha\beta}\)\p^\alpha\p^\beta h^{\mu\nu}=0\,.
\ea
We now focus on a pure cosmological constant source, $T\mn=-\lambda \eta\mn$,
and make use of a similar ansatz as previously,
\ba
h\mn&=&-\frac12 H^2x^2 \mpl\, \eta\mn\,,\\
\pi&=&\frac 12\, \kk \, x^2 \Lambda_3^3\,.
\label{anz}
\ea
The equations of motion then simplify to
\ba
\label{eqh_2}
&&\(-\frac12\mpl H^2+\sum_{n=1}^3a_n\, \kk^n \Lambda_3^{3}\)
\eta\mn=-\frac{\lambda}{6\mpl}\eta\mn \,,\\
&& H^2\(a_1+2 a_2 \kk+3 a_3 \kk^2\)=0\,.\label{eqpi_2}
\ea
As we will see below, this system of equations admits  two branches of solutions,
a ``degravitating" one, for which the geometry remains flat (mimicking the late-time part of the relaxation process),
and a ``de Sitter" branch which is closely related to the standard GR de Sitter solution.
We start with the degravitating branch before exploring the more usual de Sitter
solution and show that the stability of these branches depends on the free
parameters $a_{2,3}$,  as well as the magnitude of the cosmological constant.

\subsubsection{The degravitating branch}
In this formalism, it is easy to check that the geometry can remain
flat \ie $H=0$ and $g\mn\equiv \eta\mn$, despite the presence of the
cosmological constant. Such solutions are possible due to the presence
of the extra helicity-0 mode that ``carries" the source instead of the
usual metric. With $H=0$,  equation \eqref{eqpi_2} is trivially satisfied,
while  the modified Einstein equation  \eqref{eqh_2} determines the
coefficient (which we denote by $\kk_0$ here)
for the helicity-0 field in (\ref {anz}),
\ba
\label{pi0Jordan}
a_1\kk_0+ a_2 \kk_0^2+a_3 \kk_0^3=-\frac{\tilde \lambda}{6}\,,
\ea
in terms of the dimensionless quantity $\tilde \lambda=\lambda/\Lambda_3^3 \mpl$.
Notice that as long as the parameter $a_3$ is present, Eq.~\eqref{pi0Jordan} has always
at least one real root. There is therefore a flat solution for arbitrarily large
cosmological constant.

Let us now briefly comment on the stability of the flat  solution,
as this has important consequences for the relaxation mechanism behind degravitation.
We consider the field fluctuations  above the static solution,
\ba
\label{Pert_degrav}
\pi&=&\frac{1}{2}\kk_0 \Lambda^3_3\, x^2- \phi/\kappa\,, \\
T_{\mu\nu}&=&-\lambda \eta_{\mu\nu}+\tau_{\mu\nu} \,,
\ea
where $\kk_0$ is related to $\lambda$ via  \eqref{pi0Jordan}
and the coupling $\kappa$ is determined by
\ba
\kappa=2(a_1+2 a_2 \kk_0 +3 a_3 \kk_0^2)\,.
\ea
To the leading order, the
action for these fluctuations  is then simply given by
\ba
\mathcal{L}^{(2)}=-\frac12 h^{\mu\nu}\mathcal{E}^{\alpha\beta}\mn
h_{\alpha\beta}-\frac 12  h^{\mu\nu}X^{(1)}\mn[\Phi]+\frac{1}{\mpl}h^{\mu\nu}\tau\mn\,,
\ea
with $\Phi\mn=\partial_\mu\partial_\nu\phi$.
The stability of this theory is better understood when working in the
Einstein frame where the helicity-0 and -2 modes decouple.
This is achieved by performing the change of variable,
\ba
h\mn=\bar h\mn+ \phi \eta\mn \,,
\ea
which brings the action to the following form
\ba
\mathcal{L}^{(2)}=-\frac12 \bar
h^{\mu\nu}\mathcal{E}^{\alpha\beta}\mn
\bar h_{\alpha\beta}+\frac 32 \phi\Box \phi+\frac{1}{\mpl}
\(\bar h^{\mu\nu}+\phi\, \eta^{\mu\nu}\)\tau\mn\,.
\ea
Stability of the static solution is therefore manifest for any region of the
parameter space for which $\kappa$ is real and does not vanish.
As already mentioned, if $a_3\ne 0$ there is always a real solution to
\eqref{pi0Jordan},  which is therefore  stable for $\kappa\ne 0$.
Furthermore, direct calculations to the $6^{\rm th}$ order show that the helicity-1
fluctuations will have
a positive kinetic term as long as $\kappa / (q_0-1)> 0$. This suggests the presence of a flat
late-time attractor solution for degravitation. The special case $a_3=0$
is discussed separately below.

\subsubsection{de Sitter branch}
In the presence of a cosmological constant, the field equations \eqref{eqh_2} and
\eqref{eqpi_2} also admit a second branch of solutions; these connect with
the self-accelerating  branch presented in section~\ref{SelfAcc}, and we refer
to them as the de Sitter solutions. The parameters for these solutions
should satisfy
\ba
&&a_1+2 a_2\kk_{\rm dS}+3a_3\kk_{\rm dS}^2=0\,,\\
&&H_{\rm dS}^2=\frac{\lambda}{3\mpl^2}+
\frac{2 \Lambda_3^3}{\mpl}\(a_1 \kk_{\rm dS}+a_2  \kk_{\rm dS}^2 +a_3  \kk_{\rm dS}^3\)\,.
\ea
This solution is closer to the usual GR de Sitter configuration
and only exists if $a_2^2\ge 3 a_1 a_3$.
The stability of this solution can be analyzed as previously
by looking at fluctuations around this background configuration,
\ba
\pi&=&\frac{1}{2}\kk_{\rm dS}\, \Lambda^3_3\, x^2+\phi \,,\\
h\mn&=&-\frac 12 H^2_{\rm dS}\, x^2 \, \eta\mn+\chi\mn \,,\\
T_{\mu\nu}&=&-\lambda \eta_{\mu\nu}+\tau_{\mu\nu}\,.
\ea
To second order in fluctuations, the resulting action is then of the form
\ba
\mathcal{L}^{(2)}=-\frac12 \chi^{\mu\nu}
\mathcal{E}^{\alpha\beta}\mn \chi_{\alpha\beta}+\frac{6H_{\rm dS}^2\mpl}
{\Lambda_3^3}(a_2+3 a_3 \kk_{\rm dS})
\phi\Box \phi+\frac{1}{\mpl}\chi^{\mu\nu}\tau\mn\,.
\ea
It is interesting to point out  again that the helicity-0 fluctuation $\phi$ then decouples
from matter sources at quadratic order (however the coupling reappears
at the cubic order). Stability of this solution is therefore ensured if
the parameters satisfy one of the following three constrains, (setting $a_1=-1/2$ and $\tilde \lambda>0$)
\ba
\label{bounddS1}
a_2<0\hspace{10pt}{\rm and}\hspace{10pt}
-\frac{2a_2^2}{3}\le a_3<\frac{1-3a_2 \tilde
\lambda-(1-2a_2 \tilde \lambda)^{3/2}}{3\tilde\lambda^2}\,,
\ea
or
\ba
\label{bounddS2}
a_2 <\frac1{2\tilde{\lambda}}\hspace{10pt}{\rm and}\hspace{10pt}
a_3>\frac{1-3a_2 \tilde \lambda+
(1-2a_2 \tilde \lambda)^{3/2}}{3\tilde\lambda^2}\,,
\ea
or
\ba
\label{bounddS3}
a_2 \ge\frac1{2\tilde{\lambda}}\hspace{10pt}{\rm and}\hspace{10pt}
a_3>-\frac 23 a_2^2\,.
\ea
These are consistent with the results \eqref{bounds} found for the self-accelerating
solution in the absence of a cosmological constant. Moreover, the requirement of
stability of helicity-1 fluctuations does not impose further bounds on the parameters
(see, discussions at the end of section 3.2). Notice here that in the presence of
a cosmological constant, the accelerating solution can be stable even when $a_3=0$.
This branch of solutions therefore connects with the usual de Sitter one of GR.

\subsubsection{Diagonalizable action}

In section \ref{SelfAcc} we have emphasized the
importance of the contribution of $X^{(3)}\mn$
for the stability of the self-accelerating solution.
However, in the presence of a nonzero cosmological constant,
this contribution is not {\it a priori} essential for stability of either
the degravitating or the  de Sitter branches.
Furthermore,  since the helicity-0 and -2 modes can  be diagonalized at the nonlinear level
when $a_3=0$, as was explicitly shown in \cite{CdRGG}, we will study this special
case separately below. In particular,  we will show that it leads to certain special
bounds both in the degravitating and de Sitter branches of solution.

\vspace{0.1cm}

{\it Stability:} To start with, when $a_3=0$,  the degravitating solution only exists if
\ba
\label{constrain}
2a_2\tilde \lambda<3 a_1^2\,.
\ea
This bound, along with the stability condition for the helicity-1 ${4a_2 q_0 -1\over q_0-1} >0$,
then also ensures the absence of ghost-like instabilities around the
degravitating solution. Assuming that the parameters $a_{1,2}=\mathcal{O}(1)$ take some
natural values then the situation $a_2>0$ implies a severe constraint
on the value of the vacuum energy that can be degravitated.
This is similar  to the bound in the non-linear realization of massive gravity
\cite{CdR},  as well  as in codimension-two deficit angle solutions,
$\lambda\lesssim m^2 \mpl^2$.
The situation $a_2<0$ on the other hand allows for an arbitrarily large CC.

On the other hand, the bound $a_2^2\ge3 a_1 a_3$ for the existence of the
de Sitter solution is always satisfied if $a_3=0$. However,
the constraints on the parameters \eqref{bounddS1} - \eqref{bounddS3}
which guarantee the absence of ghosts on the de Sitter branch imply
that
\ba
2 a_2 \tilde \lambda > 3 a_1^2\,.
\ea
In this specific case then, we infer that when the Sitter solution is stable,
the degravitating branch does not exist, and when the degravitating branch exists
the de Sitter solution is unstable.  Therefore, at each point in the
parameter space there is only one,  out of these two solutions,  that makes sense.
In the more general case where $a_3\ne 0$ the situation is however much more
subtle and it might be possible to find parameters for which both branches exist and
are stable simultaneously.

\vspace{0.1cm}

{\it Einstein's frame:} Finally, to understand how this degravitating branch connects
with the arguments in \cite{Dvali:2007kt} and how it relates with Galileon theories, let us
now work instead in the Einstein frame, where the helicity-2 and -0
modes are diagonalized (which is possible as long as $a_3=0$).
The transition to Einstein's frame is performed by the change of
variable \cite{CdRGG, deRham:2010eu}
\ba
h\mn=\bar h\mn -2 a_1 \pi \eta\mn+\frac{2a_2}{\Lambda_3^3} \p_\mu\pi \p_\nu \pi\,,
\ea
such that the action takes the form
\ba
\mathcal{L}&=&-\frac{1}{2}\bar h^{\mu\nu}(\mathcal{E}\bar h)\mn+6 a_1^2 \pi \Box \pi-
\frac{6 a_2 a_1}{\Lambda_3^3}(\p \pi)^2 [\Pi]
+\frac{2a_2^2}{\Lambda_3^6}(\p \pi)^2\([\Pi^2]-[\Pi]^2\)\nn\\
&+&\frac1{\mpl} \(\bar h_{\mu\nu} -2a_1\pi \eta\mn+\frac{2a_2}{\Lambda_3^3}
\p_\mu\pi \p_\nu \pi\)T^{\mu\nu}
\,,
\ea
and the structure of the Galileon becomes manifest. Notice however, that the
coefficients of the different Galileon interactions are not arbitrary.
Furthermore,  the coupling to matter includes terms of the form
$\p_\mu\pi \p_\nu \pi T^{\mu\nu}$, absent in the original Galileon
formalism \cite{nic}. Both of these distinctions
play a crucial role in screening the cosmological constant --
the task which was thought impossible in the
original Galileon theory. Here,  however,  as long as the
bound \eqref{constrain} is satisfied, the solution for $\pi$  reads
\ba
\label{pi0Einstein}
\pi=\frac{1}{2}\, \kk_{0}\, \Lambda^3_3 \, x^2 \hspace{20pt}{\rm with}\hspace{20pt}
a_1 \kk_0+a_2\kk_0^2=-\frac{\tilde \lambda}{6},
\ea
while the helicity-2 mode $\bar h\mn$ now takes the form
\ba
\bar h\mn=\(\frac\xi2-\frac \lambda{6\mpl}\)x^2\eta\mn+\xi \, x_\mu x_\nu\,,
\ea
with  $\xi$ being an arbitrary gauge freedom parameter.
Fixing $\xi=-2a_2 \kk_0^2 \Lambda_3^3$, the physical metric is then manifestly flat:
\ba
g\mn&=&\eta\mn+\frac{1}{\mpl}\(\bar h\mn -2a_1\pi
\eta\mn+\frac{2a_2}{\Lambda_3^3} \p_\mu\pi \p_\nu \pi\)\nn\\
&=&\eta\mn-\frac{\lambda_3^3}{\mpl}\(a_1 \kk_0+a_2\kk_0^2+
\frac{\tilde \lambda}{6}\)x^2 \eta\mn+\frac{1}{\mpl}
\(\xi+2a_2\kk_0^2\Lambda_3^3\)x_\mu x_\nu\nn\\
&\equiv& \eta\mn\,.
\ea
To reiterate, the specific nonlinear coupling to matter that naturally arises in the
ghostless theory of massive gravity is essential for the screening mechanism to work.
This allows us to understand why neither DGP nor an ordinary Galileon theory
are capable of achieving degravitation.


\subsection{Phenomenology}
Let us now focus on the phenomenology of the degravitating solution.
This mechanism relies crucially on the extra helicity-0 mode in the massive graviton. However tests of gravity severely
constrain the presence of additional scalar degrees of freedom. As is well known in theories of massive gravity,
the helicity-0 mode can evade fifth force constrains  in the vicinity of matter if the helicity-0 mode interactions
are important enough to freeze out the field fluctuations, \cite{Arkady}.

Around the degravitating solution, the scale for helicity-0 interactions are no longer governed by the parameter
$\Lambda_3$, but rather by the scale determined by the cosmological constant
$\tilde \Lambda_3\sim (\lambda/\mpl)^{1/3}$. To see this, let us pursue the analysis
of the fluctuations around the degravitating branch \eqref{Pert_degrav} and keep
the higher order interactions. The resulting Lagrangian is then
\ba
\mathcal{L}^{(2)}=-\frac12 h^{\mu\nu}\mathcal{E}^{\alpha\beta}\mn h_{\alpha\beta}-\frac 12  h^{\mu\nu}\(X^{(1)}\mn[\Phi]+\frac{\tilde a_2}{\tilde \Lambda^3}X^{(2)}\mn[\Phi]
+\frac{\tilde a_3}{\tilde \Lambda^6}X^{(3)}\mn[\Phi]\)
+\frac{1}{\mpl}h^{\mu\nu}\tau\mn\,,
\ea
with
\ba
\frac{\tilde a_2}{\tilde \Lambda^3}=-2\frac{a_2+3a_3\kk_0}{\Lambda_3^3 \kappa^2}
\sim \frac{\mpl}{\lambda}\,,
\hspace{10pt}{\rm and} \hspace{10pt}
\frac{\tilde a_3}{\tilde \Lambda^6}=-\frac{2a_3}
{\Lambda_3^6 \kappa^3}\sim \(\frac{\mpl}{\lambda}\)^2\,,
\ea
assuming $a_{2,3}\sim \mathcal{O}(1)$.
To evade fifth force constrains within the solar system, the scale $\tilde \Lambda$
should therefore be small enough to allow for the nonlinear
interactions to dominate  over the quadratic contribution
and enable the Vainshtein mechanism.  In the DGP model this
typically imposes the constraint,  ${ \tilde \Lambda^3/\mpl} \lesssim (10^{-33}~{\rm eV})^2$,
while this value can be pushed by a few orders of magnitude in the presence
of Galileon interactions, \cite{nic,Burrage:2010rs}.
Therefore, the allowed value of vacuum energy  that
can be screened without being in conflict with observations is fairly low,
of the order of  $(10^{-3}~{\rm eV})^4$ or so.

Notice that this maximal cosmological constant is at least of the same order
of magnitude, if not better, than the tension that can be carried
by a codimension-2 brane embedded in six dimension with a Planck scale $M_6$.
In this scenario, the maximal tension is of the order of $\lambda<2\pi M_6^4$.
From a four-dimensional point of view, this model with the brane-induced Einstein-Hilbert
term looks like a theory of massive gravity with a graviton mass $m^2\sim M_6^4/\mpl^2$.
Phenomenology imposes the graviton mass to be $m\lesssim 10^{-33}$eV, which therefore
implies the upper bound of the brane tension,  $\lambda \lesssim (10^{-3}~{\rm eV})^4$.

The above constraint on the vacuum energy that can be degravitated makes
the present framework not viable phenomenologically for
solving the old cosmological constant problem. There may be
a way out of this setback though: As mentioned in the first section,
one may  envisage a cosmological
scenario in which the neutralization of vacuum energy takes place before
the Universe enters the epoch for which the Vainshtein mechanism
is absolutely necessary to suppress the helicity-0 fluctuations.
Such an epoch should certainly be before  the radiation domination.
During that epoch, however, the cosmological evolution should reset
itself --perhaps via some sort of
phase transition -- to continue subsequent evolution along the other  branch
of the solutions that exhibits the standard early behavior followed by the
self-acceleration,  found in the present work. This scheme would have  to
address the cosmological instabilities discussed in Refs. \cite {Luca,Stefan}.
Moreover, the viability of such a scenario
would depend on properties of the degravitating solution in the full theory --which are
not known. Therefore, we do not rely on this possibility.

Nevertheless, there are certain important virtues to
the degravitating solution with the low value of the degravitated CC.
This is an example of high  importance in understanding how S. Weinberg's
no-go theorem can be evaded in principle. As already emphasized in
\cite{deRham:2007xp,deRham:2009wb,Padilla:2010de,Padilla:2010tj}, such  mechanisms evade the
no-go theorem by employing a field which explicitly breaks Poincar\'e invariance in its vacuum
configuration $\pi \sim x^2$, while keeping the physics insensitive to this breaking.
Indeed, physical observables are only sensitive to $\Pi\mn=\p_\mu\p_\nu \pi$ which is clearly
Poincar\'e invariant, while the configuration of the $\pi$ field itself has no direct physical bearing.
This is built in the specific Galileon symmetry of the theory, and is a consequence of the fact
that $\pi$ is not an arbitrary scalar field but rather descends   as the helicity-0 mode of the massive
graviton.  More precisely, under a Poincar\'e transformation, $x^\mu\to \Lambda^\mu_{\, \nu} x^\nu+a^\mu$,
the configuration for $\pi$ transforms as $x^2\to x^2 + v_\mu x^\mu+c$,
with $v_\mu=2a_\nu\Lambda^\nu_{\, \mu}$ and $c=a^2$ which is precisely the Galileon transformation
for $\pi$ under which the action is invariant. In other words the  Poincar\'e symmetry is still
realized up to a Galilean transformation (or, there is a diagonal subgroup of
Poincar\'e and internal ``galilean'' groups that remains unbroken by the VEV of the $\pi$ field).

Thus, we have presented here the crucial steps towards a non-linear realization of degravitation
within the context of massive gravity,  and this,  without introducing any ghosts (at least in the decoupling limit).
The arguments presented here only rely on the decoupling limit and it is reasonable to doubt their validity beyond that regime. Fortunately, non-linear theories of massive gravity have been explicitly
formulated in \cite{GG,CdR}, and static solutions in the fully non-linear regime have been presented in \cite{CdR}. The absence of the ghost in theories of massive gravity requires the presence of additional symmetry projecting out the usual Boulware-Deser ghost, which can typically be thought as inherited from a higher dimensional fundamental theory. It is therefore only natural to investigate massive gravity as arising in braneworld models embedded in (spurious) extra dimensions. The static solutions presented so far then embrace a much more physical meaning, where the quantity $\Pi\mn$ plays the role of the extrinsic curvature on the brane, describing the brane position along the extra dimension(s). The fact that our model allows for flat solutions while carrying the cosmological constant with $\Pi\mn$ suggests that such models could be understood as flat branes embedded in extra dimensions, similarly as in \cite{GG,CdR}, \cite{Gabadadze:2003ck,deRham:2007xp} and \cite{deRham:2009wb}.

\vspace{0.1cm}

Some interesting work in Refs.~\cite{Padilla:2010tj} appeared during the completion of this manuscript.
These have certain overlaps with the ideas of section \ref{Degravitation} of the present work.
In particular,  Refs.~\cite{Padilla:2010tj}  emphasize  the role of Galileon
fields in the context of degravitation. These works  differ,   however,  in  several aspects
from the present one. In particular, Ref.~\cite{Padilla:2010tj} relies on the existence of
two Galileon fields, as would arise in models with two extra dimensions, \cite{deRham:2007xp},
whilst our model explores the degravitating solutions with a unique extra helicity-0 mode which naturally
arises in the 4D theory of massive gravity. Our mechanism  is possible thanks to the very specific coupling to
matter that arises in a ghostless theory of massive gravity, and differ from the standard Galileon coupling.

\vspace{10pt}

\noindent {\bf Acknowledgements:}
We wish to thank Giga Chkareuli, Sergei Dubovsky, Gia Dvali,
Justin Khoury,  Massimo Porrati and Andrew Tolley for useful discussions.
The work of CdR and LH is funded by the SNF. The work of GG was supported
by NSF grant PHY-0758032. DP is supported by the Mark Leslie
Graduate Assistantship at NYU.


\begin{thebibliography}{99}


\bibitem{acc}
  A.~G.~Riess {\it et al.}  [Supernova Search Team Collaboration],
  Astron.\ J.\  {\bf 116}, 1009 (1998)
  [arXiv:astro-ph/9805201];
  S.~Perlmutter {\it et al.}  [Supernova Cosmology Project Collaboration],
  Astrophys.\ J.\  {\bf 517}, 565 (1999)
  [arXiv:astro-ph/9812133].


\bibitem{Weinberg:1988cp}
  S.~Weinberg,
  Rev.\ Mod.\ Phys.\  {\bf 61}, 1 (1989).



\bibitem{Polchinski}
  J.~Polchinski,
{\it Talk given at 23rd Solvay Conference in Physics: The Quantum Structure of Space and Time,
Brussels, Belgium, 1-3 Dec 2005. Published in *Brussels 2005, The quantum structure of space and
time* 216-236};    arXiv:hep-th/0603249.


\bibitem{DGP}
  G.~R.~Dvali, G.~Gabadadze and M.~Porrati,
  Phys.\ Lett.\  B {\bf 485}, 208 (2000)
  [arXiv:hep-th/0005016].


\bibitem{Cedric}
  C.~Deffayet,
  Phys.\ Lett.\  B {\bf 502}, 199 (2001)
  [arXiv:hep-th/0010186].

\bibitem{DDG}
  C.~Deffayet, G.~R.~Dvali and G.~Gabadadze,
  Phys.\ Rev.\  D {\bf 65}, 044023 (2002)
  [arXiv:astro-ph/0105068].

\bibitem{Ratt}
  M.~A.~Luty, M.~Porrati and R.~Rattazzi,
  JHEP {\bf 0309}, 029 (2003)
  [arXiv:hep-th/0303116].

\bibitem{Koyama}
  D.~Gorbunov, K.~Koyama and S.~Sibiryakov,
  Phys.\ Rev.\  D {\bf 73}, 044016 (2006)
  [arXiv:hep-th/0512097].

\bibitem{Tanaka}
  K.~Izumi, K.~Koyama and T.~Tanaka,
  JHEP {\bf 0704}, 053 (2007)
  [arXiv:hep-th/0610282].

\bibitem{DGI}
  C.~Deffayet, G.~Gabadadze and A.~Iglesias,
  JCAP {\bf 0608}, 012 (2006)
  [arXiv:hep-th/0607099].

\bibitem{DW}
  G.~Dvali, G.~Gabadadze, O.~Pujolas and R.~Rahman,
  Phys.\ Rev.\  D {\bf 75}, 124013 (2007)
  [arXiv:hep-th/0612016].

\bibitem{GGcargese}
G.~Gabadadze,
  Nucl.\ Phys.\ Proc.\ Suppl.\  {\bf 171}, 88 (2007)
  [arXiv:0705.1929 [hep-th]].


\bibitem{Pujolas}
  K.~Izumi, K.~Koyama, O.~Pujolas and T.~Tanaka,
  Phys.\ Rev.\  D {\bf 76}, 104041 (2007)
  [arXiv:0706.1980 [hep-th]].

\bibitem{deRham:2006pe}
  C.~de Rham and A.~J.~Tolley,
  JCAP {\bf 0607}, 004 (2006)
  [arXiv:hep-th/0605122].


\bibitem{nic}
  A.~Nicolis, R.~Rattazzi and E.~Trincherini,
  Phys.\ Rev.\  D {\bf 79}, 064036 (2009)
  [arXiv:0811.2197 [hep-th]].


  \bibitem{CdRGG}
  C.~de Rham and G.~Gabadadze,
  Phys.\ Rev.\  D {\bf 82}, 044020 (2010)
  [arXiv:1007.0443 [hep-th]].


\bibitem{Roman}
A.~Lue, R.~Scoccimarro and G.~D.~Starkman,
  Phys.\ Rev.\  D {\bf 69}, 124015 (2004)
  [arXiv:astro-ph/0401515]; \\
R.~Scoccimarro,
  Phys.\ Rev.\  D {\bf 80}, 104006 (2009)
  [arXiv:0906.4545 [astro-ph.CO]];\\
K.~C.~Chan and R.~Scoccimarro,
  Phys.\ Rev.\  D {\bf 80}, 104005 (2009)
  [arXiv:0906.4548 [astro-ph.CO]].




\bibitem{khoury}
  N.~Afshordi, G.~Geshnizjani and J.~Khoury,
  JCAP {\bf 0908}, 030 (2009)
  [arXiv:0812.2244 [astro-ph]].

\bibitem{Arkady}
  A.~I.~Vainshtein,
  Phys.\ Lett.\  B {\bf 39}, 393 (1972).

 \bibitem{DDGV}
  C.~Deffayet, G.~R.~Dvali, G.~Gabadadze and A.~I.~Vainshtein,
  Phys.\ Rev.\  D {\bf 65}, 044026 (2002)
  [arXiv:hep-th/0106001].

\bibitem{DGZ}
  G.~Dvali, A.~Gruzinov and M.~Zaldarriaga,
  Phys.\ Rev.\  D {\bf 68}, 024012 (2003)
  [arXiv:hep-ph/0212069].

\bibitem{Lue}
A.~Lue and G.~Starkman,
  Phys.\ Rev.\  D {\bf 67}, 064002 (2003)
  [arXiv:astro-ph/0212083].

\bibitem{Appolo}
T.W.~Murphy, Jr., E.G.~Adelberger, J.D.~Strasburg, and
C.W.~Stubbs,
"APOLLO: \text{Multiplexed Lunar Laser
Ranging", at
http://physics.ucsd.edu/} \\ \text{ ~tmurphy/apollo/doc/multiplex.pdf}


\bibitem{Babichev}
E.~Babichev, C.~Deffayet and R.~Ziour,
  Phys.\ Rev.\ Lett.\  {\bf 103}, 201102 (2009)
  [arXiv:0907.4103 [gr-qc]]; \\
 E.~Babichev, C.~Deffayet and R.~Ziour,
  arXiv:1007.4506 [gr-qc].



\bibitem{DGS}
G.~Dvali, G.~Gabadadze and M.~Shifman,
{\it Published in *Minneapolis 2002, Continuous advances in QCD* 566-581};
arXiv:hep-th/0208096;\\
G.~Dvali, G.~Gabadadze and M.~Shifman,
  Phys.\ Rev.\  D {\bf 67}, 044020 (2003)
  [arXiv:hep-th/0202174];\\


\bibitem{ArkaniHamed:2002fu}
  N.~Arkani-Hamed, S.~Dimopoulos, G.~Dvali and G.~Gabadadze,
  arXiv:hep-th/0209227.


\bibitem{Dvali:2007kt}
  G.~Dvali, S.~Hofmann and J.~Khoury,
  Phys.\ Rev.\  D {\bf 76}, 084006 (2007)
  [arXiv:hep-th/0703027].



\bibitem{FP}
  M.~Fierz and W.~Pauli,
  Proc.\ Roy.\ Soc.\ Lond.\  A {\bf 173}, 211 (1939).


\bibitem{Nieu}
P.~Van Nieuwenhuizen,
  Nucl.\ Phys.\  B {\bf 60}, 478 (1973).


\bibitem{BD}
  D.~G.~Boulware and S.~Deser,
  Phys.\ Rev.\  D {\bf 6}, 3368 (1972).



\bibitem{GGruzinov}
  G.~Gabadadze and A.~Gruzinov,
  Phys.\ Rev.\  D {\bf 72}, 124007 (2005)
  [arXiv:hep-th/0312074].

\bibitem{AGS}
  N.~Arkani-Hamed, H.~Georgi and M.~D.~Schwartz,
  Annals Phys.\  {\bf 305}, 96 (2003).

\bibitem{Creminelli1}
  P.~Creminelli, A.~Nicolis, M.~Papucci and E.~Trincherini,
  JHEP {\bf 0509}, 003 (2005).

  \bibitem{DeffayetRombouts}
  C.~Deffayet and J.~W.~Rombouts,
  Phys.\ Rev.\  D {\bf 72}, 044003 (2005)
  [arXiv:gr-qc/0505134].



\bibitem{deRham:2010gu}
 C.~de Rham and G.~Gabadadze,
  Phys.\ Lett.\  B {\bf 693}, 334 (2010)
  [arXiv:1006.4367 [hep-th]].


\bibitem{GG}
  G.~Gabadadze,
  Phys.\ Lett.\  B {\bf 681}, 89 (2009)
  [arXiv:0908.1112 [hep-th]].


\bibitem{CdR}
  C.~de Rham,
  Phys.\ Lett.\  B {\bf 688}, 137 (2010)
  [arXiv:0910.5474 [hep-th]].



\bibitem{deRham:2010kj}
  C.~de Rham, G.~Gabadadze, A.~J.~Tolley,
  [arXiv:1011.1232 [hep-th]].




\bibitem{GGIglesias}
G.~Gabadadze and A.~Iglesias,
  Phys.\ Lett.\  B {\bf 639}, 88 (2006)
  [arXiv:hep-th/0603199].

\bibitem{Gabadadze:2003ck}
  G.~Gabadadze and M.~Shifman,
  Phys.\ Rev.\  D {\bf 69}, 124032 (2004)
  [arXiv:hep-th/0312289].

\bibitem{deRham:2007xp}
  C.~de Rham, G.~Dvali, S.~Hofmann, J.~Khoury, O.~Pujolas, M.~Redi and A.~J.~Tolley,
  Phys.\ Rev.\ Lett.\  {\bf 100}, 251603 (2008)
  [arXiv:0711.2072 [hep-th]];


  C.~de Rham, S.~Hofmann, J.~Khoury and A.~J.~Tolley,
  JCAP {\bf 0802}, 011 (2008)
  [arXiv:0712.2821 [hep-th]];

  C.~de Rham,
  Can.\ J.\ Phys.\  {\bf 87}, 201 (2009)
  [arXiv:0810.0269 [hep-th]].

\bibitem{deRham:2009wb}
  C.~de Rham, J.~Khoury and A.~J.~Tolley,
  Phys.\ Rev.\ Lett.\  {\bf 103}, 161601 (2009)
  [arXiv:0907.0473 [hep-th]].

\bibitem{Padilla:2010de}
  A.~Padilla, P.~M.~Saffin and S.~Y.~Zhou,
  arXiv:1007.5424 [hep-th].

\bibitem{Padilla:2010tj}
  A.~Padilla, P.~M.~Saffin and S.~Y.~Zhou,
  arXiv:1008.3312 [hep-th].

\bibitem{deRham:2010eu}
  C.~de Rham and A.~J.~Tolley,
  JCAP {\bf 1005}, 015 (2010)
  [arXiv:1003.5917 [hep-th]].


\bibitem{Burrage:2010rs}
  C.~Burrage and D.~Seery,
  JCAP {\bf 1008}, 011 (2010)
  [arXiv:1005.1927 [astro-ph.CO]].


\bibitem{Luca}L.~Grisa and L.~Sorbo,
  Phys.\ Lett.\  B {\bf 686}, 273 (2010)
  [arXiv:0905.3391 [hep-th]].


\bibitem{Stefan} F.~Berkhahn, D.~D.~Dietrich and S.~Hofmann,
  arXiv:1008.0644 [hep-th].

\end{thebibliography}
\end{document}